# SS433 Micro-quasar Jet and the TeV Resurgence beam

Daniele Fargion[1][2] *

December 11, 2024

[1] Physics Department, Rome University1,Sapienza, Pl.A.Moro 2,Rome,Italy
[2] Mediterranian Institute of Fundamental Physics, Marino, Rome, Italy

**Abstract**

The understanding of micro-quasars in our galaxy is one of the frontiers of high-energy astrophysics. Their models are based on a capturing mass Black Hole (BH) with a nearby spiraling binary companion star. The companion star mass feeds the accretion disk around the BH ,. This energy also fuels an orthogonal precessing X-gamma jets. The spiral precessing tail of such micro-quasars, as the SS433 system, is due to an ultra-relativistic jet, spraying nucleons and electrons at relativistic speeds. The up-down jet is observable in radio, X, gamma spectra. Its long spirals are spread and diluted within a light-year distance. The source is inside the W50 supernova remnant nebula , whose asymmetry reflects the past and present role of the SS433 jet. Very recently HESS, HAWC and LHAASO discovered , surprisingly at a much far disconnected distance from the SS433, the resurgence of a twin gamma beam tail. Nearly 75 years light distance far away the same inner jet source. The recent "standard model" is based on an accelerating shock wave which re-accelerates, the resurgence of a PeV nucleon beam and its TeV secondaries. The surprising re-collimation of this TeV beam jet is difficult to accept, in the assumption of a planar-like Fermi shock wave model. Here we discuss an alternative framework based on known high energy nuclear physics, capable to simultaneously explaining both the disconnected and the aligned hard TeV jet appearance. Several consequences , that might also be able to validate the model, are considered .

*Daniele.Fargion@uniroma1.it

# 1 Introduction: SS433 and its separated TeV beam

The microquasars are binary systems where a neutron star (NS), of a few or tens solar masses Black Hole (BH), are bounded by gravity with an orbiting star of a comparable (or larger) mass, while capturing the star mass by tidal forces. While such a tail of mass is collapsing onto the NS ( or the BH), the same mass feeds an accreting disk around the NS or BH. This spinning disk usually induces asymmetric charged flows and consequently huge currents that are creating huge toroidal magnetic field. Such magnetic fields may experience fast time variability. This induces also extreme electric spiral fields that accelerate and rotate, along the disk boundary, the free charges . They spin along a twin disk ring, above and below the same accretion disk. These ultra-high-energy charges are bounded, by narrow Larmor radius, spiraling along the accretion disk , accumulating on the poles. At the end, these relativistic particles are forced to be ejected vertically, aligned along the NS or BH , North or South poles axis. Ideal jet accelerator are formed at their maximal fields, near the NS or BH accretion disks. Magnetic field vibrations by their sudden shrinkage and expansions , led to the up-down charge ejection, in collimated pair jet. Tidal forces among orbital star companion and the same accretion disk, may drive to an additional conical , precession of the same jet. These events in their earliest and in the late mass accretion phase, may feed a thin persistent, precessing jet . Its blazing may be observed rarely on axis, as the far cosmic Gamma Ray Burst (GRB) or the nearer Soft Gamma Repeaters (SGR). [1]. The magnetic field lines along the jet constrain and collimate the beam. The leptonic component, electron pairs, of this jet is leading, by synchrotron radiation and by Inverse Compton Scattering (ICS) [8], to radio, X and gamma photon. These gamma signals (MeV-GeV-TeV) had been discovered since half a century and also, very recently, in surprising details.

As the microquasar name suggest they are just a small scale (parsecs) system of a famous and larger quasar. These ones are made by million or billion solar mass BH, hidden in galactic centers, called also Active Galactic Nuclei, (AGN). They are able to eject much wider, longer and harder jet beam, even in Mega parsec sizes. As we noted, SS433 is a rare binary system containing a supergiant star that is overflowing its Roche lobe with matter onto a nearby BH. Its small size, but its nearer distance , offer an ideal test for inspection of the jet engine. LHAASO experiment, the widest array km-squre size, recently studied the system [7]. The TeV beam appearance far from the source, was discovered this year by the Cherenkov telescope array, Hess [2] and confirmed also by a large array, in Mexico, HAWC [3]. The presence and collimation at such far separated beam at far distances, is quite puzzling. Most models require surprising shock wave re-acceleration and re-collimation. We shall not discuss them in present article. Here we suggest the different possibility that a rarest , early, explosive flare episode in the micro-quasar may shine, at the same time, ultra-violet hot photons and ultra-relativistic energy (UHE) proton (or nuclei) in the jet. Their photo-nuclear scattering and interaction could form also Delta $\Delta$ resonances. The $\Delta$ decay may feed pions but also ultra-relativistic neutrons. Such a phenomena has an analogy , (in cosmic volumes and in thermal big bang radiation), taking place between UHECR (Ultra High Energy Cosmic Ray) at $6 \cdot 10^{19} - 10^{20}$ eV energy and infrared cosmic black body photon at $2.7 K^o$. The phenomena is well known under the author initials, as the "GZK cut-off" in the maximal UHECR spectra [4], and [5]. The Delta resonance at ultra-relativistic energy (UHE) decay may produce neutral and charged pion respectively with associated proton or neutron: a new collimated energetic neutron jet. The presence of such tens PeV neutron beam , its first flying in a silent path and later, by its decay, in a separated visible beta trace, is the corner stone of our model. The observed 75 years light distance of the UHE neutron decay, its consequent electron radiation , by ICS, as TeV gamma resurgence beam, is the main engine solving the SS433 puzzling , disconnected , TeV tail resurgence.

## 1.1 Pevatron; A Neutron jet and its decay distance

The distance of an UHE ( about 25 PeV) neutron beta decay in flight is nearly 75 years light . Indeed:

$$L_n = 877(E_n/m_n)s \cdot c.; L_n = 75y \cdot c(E_n/25 PeV) \qquad (1)$$

A main question arises about the observed signal in SS433 , for a TeV beam discontinuity versus an expected continuity of a cosmic-ray spectra of the jet. The observed cosmic ray spectra on Earth , up to GZK cut-off edges, decrease smoothly by a power law, without any sudden peak or cut-off discontinuity. One would imagine a neutron secondary jet spectra ruled by a continuous signature in their decay distances. Our description finds a natural, nearly mono energetic neutron jet beam. See for a visual SS433 image system description, the following Figures below.

## 2 Delta resonance by proton-photon interaction

The GZK cut-off is defined by a peculiar threshold. Only at those highest energy a proton may produce also a neutron . In analogy, the tens PeV , UHE neutron jet formed by tens PeV protons in SS433 jet, require its tuned scenario. The threshold for the meson resonance is defined, as an order of magnitude, by the center of mass-energy for the $\Delta$ formation. The event is made by an UHE proton $E_p$ scattering onto a hot thermal bath of photons $E_\gamma$, from the BH accretion disk. Their ultra-relativistic approximation is :

$$\sqrt{2 \cdot E_p \cdot E_\gamma} = m_\Delta \qquad (2)$$

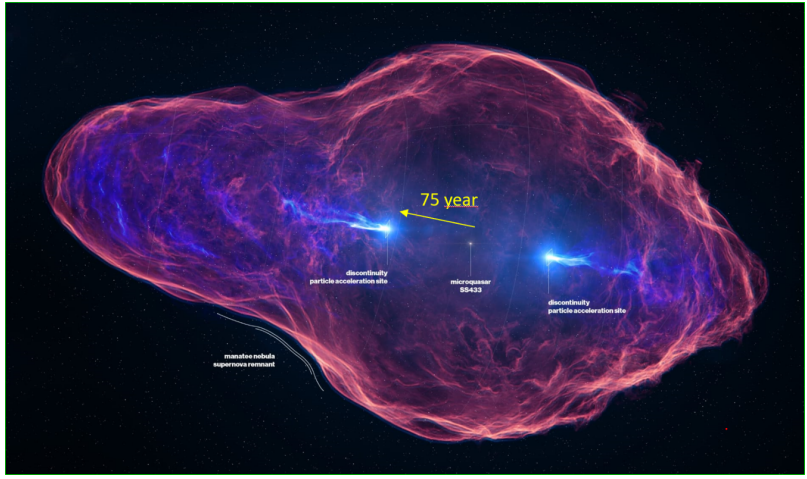

Figure 1: A simplified description of the separated TeV beam observed by Hess, HAWC and LHAASO inside the W50 nebula of a much early supernova. The resurgence of the twin TeV collimated gamma beam at $75y \cdot c$ is very puzzling.

The baryon resonant Delta mass, $m_\Delta$, has a value = 1232 MeV. This equation, for a given proton energy, defines the corresponding critical thermal photon energy $E_\gamma$. Therefore, also the tuned accretion disk temperature and its luminosity, necessary to take place for such a processes . The $\Delta$ decay may lead to a UHE proton with a neutral pion, $\pi^0$. This pion is decaying soon in a photon pair secondary. With the same rate the $\Delta$ baryon may also decay into a neutron and its pion charged companion, $\pi^+$.

The charged pion also decay soon into muon and its neutrino . Finally also the positive muon secondary , will decay into a positron and their two associated neutrino flavors. The final electromagnetic secondaries (photon pairs, electron pairs) shine and dissolve within a very near (parsec) distances from the source. They cannot reach far distances and cannot play any role at $75y \cdot c$ distance.

The tens PeV neutron beam is also created. The proton, at Pevatron energy, are bent by the galactic magnetic fields. loosing their directionality and also smearing their beam into wider spiral trajectory . Their role is negligible, because proton do not radiate much. The PeVs neutron beam, instead, may escape and fly keeping directionality, with a far decay distance, with no losses in the early flight, up to its radiating, beta decay stage.

The correlated photon energy to allow such a proton-pion, Delta resonance with a $E_p = 25$ PeV for a proton must be:

$$E_\gamma = (m_\Delta)^2/E_p = 30.35 eV/(E_p/(25 PeV)) \qquad (3)$$

One should also take care of a partial loss of energy for the final proton (or neutron) secondary, because the associated pion secondaries creation, in the Delta decay. The pions absorb part of the primary proton energy. This energy loss is just at ten percent level. Consequently, for a final 25 PeV neutron, we must assume a primary proton at higher energy, nearly of 27.5 PeV. The interacting photon energy $E_\gamma$ should be tuned to the resonance mass $m_\Delta$ by a value:

$$E_\gamma = (m_\Delta)^2/E_p = 31.6 eV/(E_p/(27.5 PeV)) = 3.6 \cdot 10^5 K^o/(E_p/(27.5 PeV)) \qquad (4)$$

This photon energy is nearly 63.6 times higher than our solar one. Let us show in the following figure the possible, present, parameters of the SS433 with their nominal comparable mass of 10 solar masses, in a circular Keplerian orbit of 13 day period ( and 162 day precession time). The star companion radius, the BH accretion disk size are shown with approximated values, while the orbit distance is tuned for their assumed 10 solar mass each. This mutual distance D may be only slightly re-calibrated, for any different binary masses, following the well known Keplerian (cubic root) law:

$$D = 146 \cdot ((M_{BH}/10M_o) + (M_{Star}/10M_o))^{1/3} s \cdot c \qquad (5)$$

## 2.1 Temperature, photon density, SS433 flare luminosity

Let us assume for sake of simplicity that the accretion disk, in the figure assumed about $0.8 s \cdot c$ size, extends a little more, with a radius comparable to our Sun or better, to its spherical total surface. This is taking place, for a disk radius $R_{disk} = \sqrt{2} \cdot R_{Sun}$. Because, as shown above, the disk flare temperature required for the resonance to arise is nearly 63 times the solar one, the corresponding peak flare luminosity $L_{Flare}$ should be (by scale Stephan Botzmann law), approximately, $(63)^4$ times larger than our Sun., or

$$L_{Flare} = 6.03 \cdot 10^{40} \cdot erg/s \qquad (6)$$

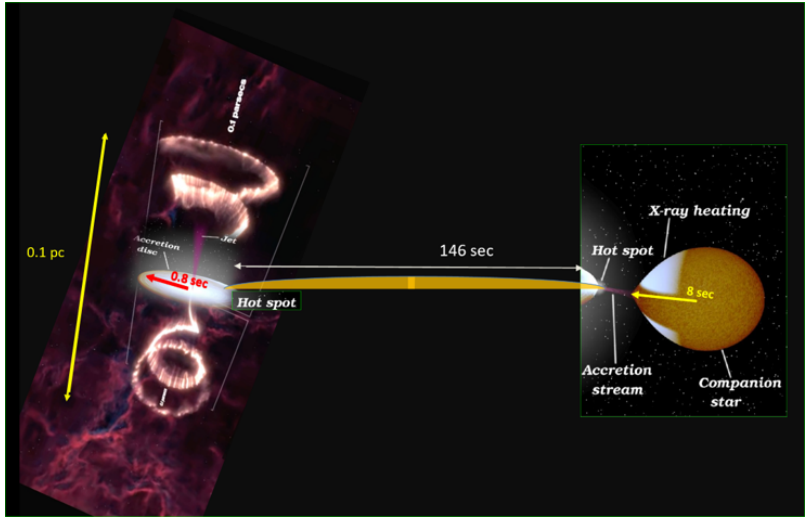

Figure 2: A simplified description of the inner SS433 binary, in approximated size, system: an accretion disk on a BH and a companion star of comparable mass, both of them, with a nominal ten solar mass, in their corresponding Keplerian circular distance. The star radius and in particular, the accretion disk, are approximated just for a comprehensive view, but they may be also a little or quite larger.

It should also remind here that this total flare power (considering the loss of energy in mass ejection and loss inside the BH) imply a much larger mass-energy rate waste, falling into the BH. We suggest an energy output (and a corresponding mass rate from the companion star to the BH) thousand times larger, as large as : $(dM/dt)_{tot} \cdot c^2 > 6.03 \cdot 10^{43} \cdot erg/s$ .

This losses are not far from the usual SS433 observed ones, but they require a much brighter luminosity, near a Nova like , $(dM/dt)_{Nova} > 10^{44} \cdot erg/s$ , explosive event. The total duration of such a huge explosion may be of few hours or even, a very few days. For such a brief duration, the SS433 precessing jet (162 day period) , its deflection, could be considered negligible, or just, to be "frozen" , pointing in a unique direction, as the observed separated TeV beam.

## 2.2 Delta resonance creation in SS433 by Nova-like event

As above estimate, the total mass in rapid capture may be as large as a part of a thousand a solar mass. Thus such a high brightening might be due to a very nearby tidal encounter of the companion star mass , or, more probably, due to the capture of a large wind or even a planet (Jupiter like) in near orbit along the companion star, suddenly falling onto the BH own gravity. Such a Nova like Burst , 75-80 years ago, might be occurring at the end of War World II, when astronomy was probably not carefully and actively observing that sky region. The same SS433 system nature, indeed, had been discovered much later, on 1977. If our hypothesis of a Nova-like flare is correct, more inspections of the past sky photos in SS433 direction , recorded $75 - 80$ years ago, may disentangle that correlated event.

As we know, the maximal , peaked, cross section $\sigma$ , for a photo-pion resonance, has a value as large as:

$$\sigma_\Delta = 500 b \tag{7}$$

Such a peaked cross-section explains the consequent, almost monochromatic, neutron beam energy. Such a hot thermal photon bath, its number density , should allow an interaction probability above the threshold. This UV temperature considered above, emitted by a black body area, an "over-Eddington" accretion disk, defines also the same photon number density $n_{Th=3.676 \cdot 10^5 K^o}$ along the jet ultra-relativistic proton propagation fly: the main interaction distance nearby the disk, $D_j$ could be comparable with the same accretion disk radius, leading to an approximated probability $P_{Photo-pion}$, for the photo-pion conversion into a Delta (and a consequent neutron beam creation) as follows:

$$P_{Photo-pion} = \sigma_\Delta \cdot n_{Th} \cdot D_j = 48 >> 1 \tag{8}$$

This large order of magnitude for a Delta resonance production, guarantees also a realistic neutron beam creation rate. Even at a far (hundred sec c distance) from the same accreting SS433 disk. The agreement between the neutron beam energy, to fit the 75 $y \cdot c$ , the corresponding flare temperature, and the associated thermal photon number density, all of them overcome the needed threshold , $P_{Photo-pion} >> 1$, for the $\Delta$ production. All the puzzle pieces are offering support to the present model thesis.

## 2.3 Larmor radius for PeV proton and TeV electrons

As we had proposed above, tens PeV neutron decay, $75yc$ far away, might become source of the observed , separated, 25 TeV gamma beam. We may inquire to the role of the twin proton secondaries of the $\Delta$ decay at comparable energies. These 25 PeV energetic proton, contrary to neutron ones, are bent. Their Larmor radius $R_p$ is constrained by the minimal galactic disk magnetic fields $B_g$ , (at least , of few micro Gauss) :

$$R_p = (E_p/25 PeV) \cdot (B_g/(3 \cdot \cdot Gauss))^{-1} \cdot 26.4 \cdot yc \tag{9}$$

These spiral radius are nearly one third of the 75 TeV beam distance. Therefore these proton components are diffused and smeared: they cannot play a role in the separated TeV jet. Also the secondary electrons at nearly 25 TeV energy , the parasite secondaries traces of the neutron beta decay, are even more confined along the neutron beta trajectory, within a much smaller spirals of radius $R_e$

$$R_e = (E_e/25 TeV) \cdot (B_g/(3 \cdot \cdot Gauss))^{-1} \cdot 0.026.4 \cdot yc \qquad (10)$$

Therefore the electron role is not able to survive hundreds years light, in absence of the suggested primary 25 PeV energetic neutron beam.

### 2.4 Neutrons by photo-nuclear Giant Dipole Resonance

Recent UHECR model imagined lightest nuclei currier being fragmented by photo-nuclear distruption, via the Giant Dipole Resonance , GDR., in concurrence with "standard model foreseeing the proton or iron UHECR courier. The GDR process, in analogy with the previous photo-pion one, is capable to lead among the fragments, also to UHE neutrons. The Virgo UHECR signal absence is a paradox for any UHECR proton courier model. The lightest nuclei model opacity toward UHECR was considered mainly to filter or hide such (onobserved) Virgo cluster presence . The GDR processes has a lower energy threshold than photo-nucleon resonance, but it is not acting on a proton or a neutron: it regards only nuclei. Here we just mention this additional opportunity windows for neutron creation and beaming jet. We also remind the possibility that such neutron process extend, from few $10^{16}$ eV to several tens EeV, above $10^{19}$ eV energy (by wider accretion disk and lower energy red-shifted photons) [9]. In this extreme scenario, we noted the very rare clustering of 4 UHECR events by AUGER and by TA array detectors in last two decades, all of them overlapping within a narrow spot of events at SS433 direction. Their few degrees collimation and their short time (decades) period of recording, might be well consistent , as discussed elsewhere [6], with such tens EeV energy neutron collimated flight . The GDR processes considered here does not need to be the most probable one for the neutron beam existence, but it must be taken in consideration.

## 3 Conclusions

A past flare in SS433, could be the source of a 25 PeV neutron jet , possibly explaining the puzzling separated twin, TeV , gamma beam at 75 years far distance. The unobservable PeV neutron flight show its presence and its beam resurgence as soon it decay at far distance and its electron may shine by synchrotron and ICS radiation. its collimation is an advantage respect the standard model of a schock wave processes . This ultra-relativistic beta decay imply also additional interesting consequences:

An early trigger explosive event, nearly 75/80 years ago, as powerfull as a Nova star burst, for a duration of few hours or few days at SS433 system shine almost in a fixed direction (not along a conical precessing jet volume). This event occurrence is around the War World II end times. Such a Nova event ( to-day observed at a rate of a dozen a year), could have been escaped detection, at those post war times . Indeed the same nature of SS433 had been discovered much later, on 1977. It could be possible and worth-full to inspect such luminosity variability, in oldest astronomical photo-plate array in that direction and at those epochs, looking for such a variability signal.

The presence of such neutron PeV separated beam in SS433, may suggest also the search in other micro-quasars system elsewhere, for such disconnected signatures. Their statistics may define a corresponding neutrino spectra rate in PeV and hundred TeV energy range,

The 25 PeV neutron beta decay and its primary 27 PeV proton-pion event, while on axis toward us , may shine both of a brightest prompt $(1-2)$ PeV gamma burst but also of a secondary (electron and muon ones) neutrino at $(0.2-0.5)$ PeV energy . Such a tuned energies are quite interesting , because they may reflect in the TeV PeV apparent energy dis-continuity, in neutrino spectra . Such spectra feature might be already hidden in the highest energy Ice-Cube neutrino data, assuming that their observed origin is really astrophysical and not , as some suggested, a charmed atmospheric noise.

## Acknowledgment and Dedication

The author thanks the kind support of Prof. M.Khlopov in discussions and in support of the article. The author dedicate the article to the memory of the victims and the hostages of the 7 october 2023 massacre , the worst evil pogrom of the century .

.